\documentclass[aps,prl,twocolumn,groupedaddress,showpacs,superscriptaddress,amssymb,amsmath,longbibliography]{revtex4-2}
\usepackage{graphicx}
\usepackage{amsmath}
\usepackage{amssymb}
\usepackage{amsfonts}
\usepackage{color}
\usepackage{bm}        
\usepackage{comment}
\usepackage{placeins}
\usepackage[mathscr]{eucal}
\usepackage[colorlinks, linkcolor=red,citecolor=blue,urlcolor=blue]{hyperref}
\usepackage[normalem]{ulem}
\usepackage[all]{hypcap}
\usepackage{multirow}
\usepackage[dvipsnames,usenames,table]{xcolor}
\usepackage{dcolumn}
\usepackage{hyperref}
\usepackage{bm}
\usepackage{epsf}
\usepackage{subfigure}
\usepackage{amsfonts}
\usepackage{epstopdf}%
\newcommand{\be}{\begin{equation}}
	\newcommand{\ee}{\end{equation}}
\newcommand{\bea}{\begin{eqnarray}}
	\newcommand{\eea}{\end{eqnarray}}
\newcommand{\me}{\mathrm{e}}
\newcommand{\ket}[1]{\vert #1 \rangle}
\newcommand{\bra}[1]{\langle #1 \vert}
\newcommand{\braket}[3]{\left\langle #1 \middle| #2 \middle| #3 \right\rangle}

\newcommand{\nn}{\nonumber}

\begin{document}
	\title{Optimal control of quantum thermal machines using machine learning} 
	\author{Ilia Khait}
	\affiliation{Department of Physics, University of Toronto, Toronto, Ontario, Canada M5S 1A7}
	\author{Juan Carrasquilla}
	\affiliation{Vector Institute, MaRS Centre, Toronto, ON, Canada}
	\affiliation{Department of Physics \& Astronomy, University of Waterloo, Waterloo, ON, Canada}
	\author{Dvira Segal}
	\affiliation{Department of Chemistry and Centre for Quantum Information and Quantum Control,
		University of Toronto, 80 Saint George St., Toronto, Ontario, M5S 3H6, Canada}
	\affiliation{Department of Physics, University of Toronto, Toronto, Ontario, Canada M5S 1A7}
	\email{dvira.segal@utoronto.ca}
	
	\date{\today}

	\begin{abstract}
		Identifying optimal thermodynamical processes has been the essence of thermodynamics since its inception.
		Here, we show that differentiable programming (DP), a machine learning (ML) tool, can be employed to optimize finite-time thermodynamical processes in a quantum thermal machine.
		We consider the paradigmatic quantum Otto engine with a time-dependent harmonic oscillator as its working fluid, and build upon shortcut-to-adiabaticity (STA) protocols. 
		We formulate the STA driving protocol as a constrained optimization task and apply DP to find optimal driving profiles for an appropriate figure of merit. Our ML scheme discovers profiles for the compression and expansion strokes that are superior to previously-suggested protocols.
		Moreover, using our ML algorithm we show that a previously-employed, intuitive energetic cost of the STA driving suffers from a fundamental flaw, which we resolve
		with an alternative construction for the cost function. 
		Our method and results demonstrate that ML is beneficial both for solving hard-constrained quantum control problems 
		and for devising and assessing their theoretical groundwork.
	\end{abstract}

	\maketitle 
	
	Many problems in physics are formulated as optimization tasks by identifying a cost function that has to be minimized. 
	Prime examples are Hamilton's principle of least action in Lagrangian mechanics~\cite{variation}, 
	Fermat's law of least time in classical optics~\cite{Lipson}, and more recently, variational algorithms in quantum computing~\cite{VQE}.
	Similarly, since its inception, thermodynamics was  concerned with performance optimization by
	identifying constrains and bounds on energy conversion processes.
	The ideal Carnot engine is designed to reach the maximal efficiency.
	However, this upper bound is theoretically obtained for arbitrarily slow, quasistatic processes, thus
	the extracted power reduces to zero.
	Quasistatic processes are described using the framework of equilibrium thermodynamics. In contrast, real thermal devices operate on finite-time cycles, and they are naturally described in terms of finite-time thermodynamics~\cite{gemmer2009quantum,finitetimeT}. This theory is 
	concerned with e.g. 
	how  the  efficiency  of thermal machines erode when heat-to-work conversion processes take place in finite-time cycles~\cite{PhysRevLett.105.150603,Esposito_2010}.
	
	Quantum thermal machines, in which e.g. quantum coherences, correlations, and quantum statistics play a decisive role
	cater fundamental understanding of thermodynamics at the nano and atomistic scale~\cite{DeffnerThermoBook,bhattacharjeeQuantumThermalMachines2020}.  
	Beyond fundamental interest, quantum thermal machines 
	promise  compact, fast, and efficient  work extraction and refrigeration schemes for quantum  devices. 
	It remains however a challenge to harness such effects and achieve a quantum advantage in thermal  machines
	~\cite{Kosloff_2014,Janet,dasQuantumenhancedFinitetimeOtto2020, PhysRevLett.122.110601,PhysRevResearch.2.043247}. 
	
	Optimizing the performance of {\it nanoscale, quantum} thermal machines is a central problem in the rapidly-emerging field of quantum thermodynamics. Techniques such as shortcut-to-adiabaticity (STA) allow the design of finite-time protocols, which reproduce the same final state of an adiabatic time evolution, yet at a price of a supplemental work on the system~\cite{PhysRevLett.104.063002,Torrontegui_2013,PhysRevA.82.053403,Muga_2010,jpca.5b06090}. 
	Much theoretical and experimental effort~\cite{Abah_2012,Santos_2015,Kosloff_2017,PhysRevB.100.035407,RevModPhys.91.045001,PhysRevApplied.13.044059,PhysRevLett.125.166802,Dupays_2021} has been put to realize and characterize these systems. Here, we focus on a specific class of STA protocols - local counterdiabatic driving (LCD), which are advantageous to the realization of quantum engines since they only require the application of local time-dependent potentials. 
	\begin{figure}[h!]
		\includegraphics[width=\columnwidth]{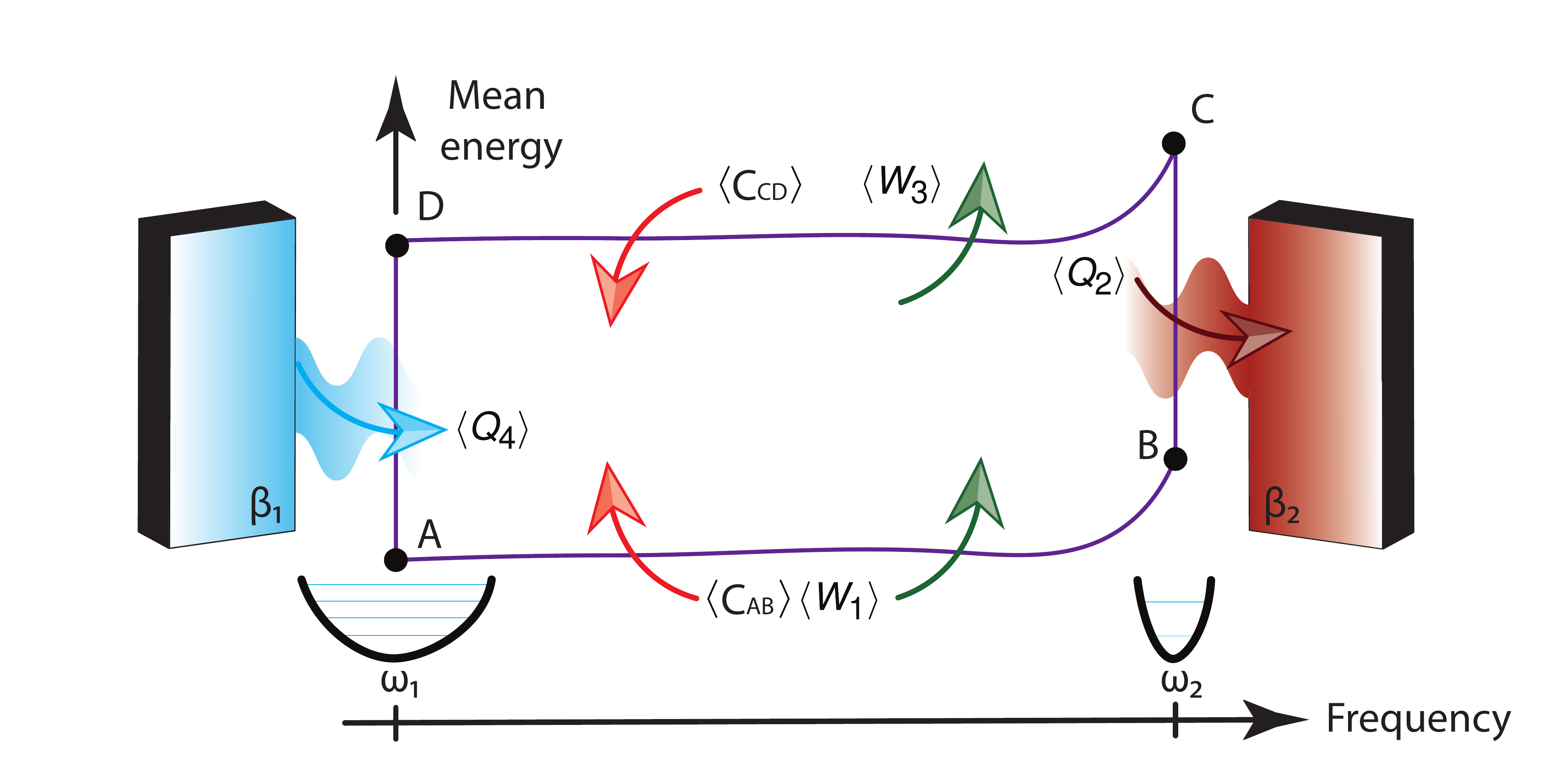} 
		\caption{Scheme of the Otto refrigerator in the energy-frequency domain. 
			A cycle includes a compression stroke (AB) of duration $\tau$, an instant isochoric stroke (BC) with the system coupled to a hot bath, an expansion stroke (CD) of duration $\tau$, and another instant isochoric stroke with a cold bath (DA). Refrigeration corresponds to the withdrawal of heat $\langle Q_4 \rangle$ from the cold bath. The energetic cost of the cycle is the sum of the work contributions $\langle W_1 \rangle$ and $\langle W_3 \rangle$, along with the energetic cost of the STA driving, $\langle C_{\rm AB}\rangle$ and $\langle C_{\rm CD}\rangle$. }
		\label{fig:ref_scheme}
	\end{figure}
	
	In this work, we harness state-of-the-art machine learning (ML) techniques to optimize the performance of quantum thermal machines. Specifically, we optimize an LCD protocol in the quantum Otto refrigerator, depicted in Fig.~\ref{fig:ref_scheme}.
	ML allows us to employ a broad, unique approach for the design, optimization and control of a large variety of classes of problems, including quantum processes~\cite{Melnikov1221,PhysRevX.8.031086,PRXQuantum.1.010301,beeler2021optimizing,PhysRevResearch.2.032051}. 
	Advances in diverse research topics, such as image recognition and natural language processing have led physicists 
	to exploit ML in quantum dynamics and many-body physics ~\cite{Juan_2020}.  
	
\begin{figure}[h!]
		\includegraphics[width=\columnwidth]{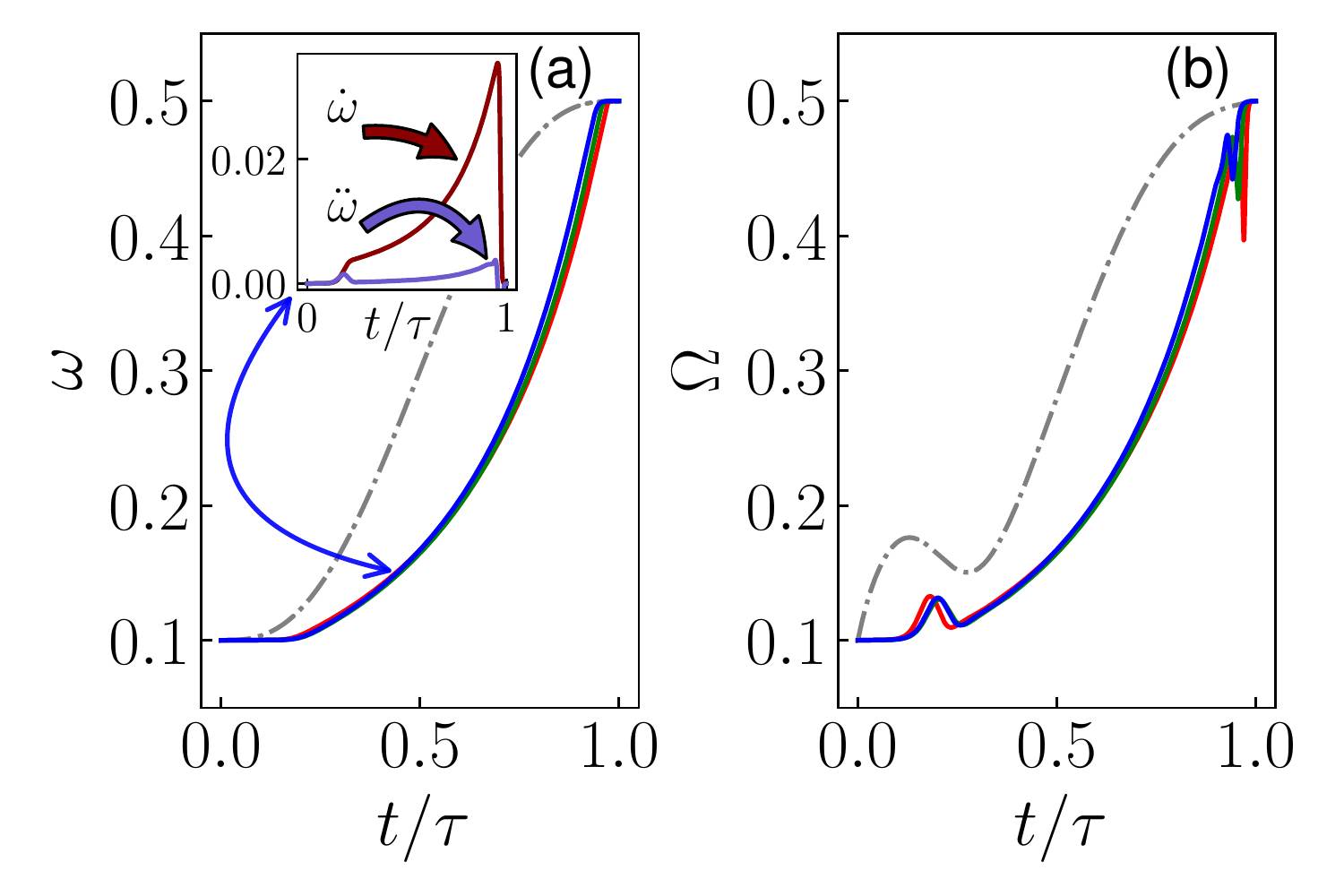} 
		\vspace{-7mm}
		\caption{(a) Examples of frequency profiles $\omega(t)$ discovered by the DP-ML scheme normalized by the compression or expansion stroke duration $\tau$. The gray dashed-dotted line displays the polynomial ansatz, Eq.~(\ref{eq:poly}). The solid lines are the neural network results. The inset displays the first and second derivatives of $\omega(t)$ for one  realization, showing compliance with the STA 
			conditions, Eq.~(\ref{eq:STAcond}). The fact that both derivatives approximately follow each other allows for the minimization of the energetic cost of the STA process (see Fig.~\ref{fig:d_tC}). (b) The function $\Omega(t)$ corresponding to the frequency of the effective counterdiabatically-driven harmonic oscillator. 
			Parameters are $\omega_1=0.1$, $\omega_2=0.5$, $\beta_1=1$ and $\beta_2=0.75$. } 
		\label{fig:omega}
	\end{figure}

	We adopt Differentiable Programming (DP)~\cite{JMLR:v18:17-468,schaferDifferentiableProgrammingMethod2020a} to find optimal refrigeration schemes for the quantum Otto cycle under STA conditions. 
	From a reinforcement learning (RL) perspective, in this scheme an agent plays a ``game", where the time-dependent frequency $\omega(t)$ of the harmonic oscillator acting as the working medium of the refrigerator can be varied in the time interval $t \in [0, \tau]$. 
	For each attempted strategy, $\omega(t)$, the agent receives a reward  designed to minimize the energetic cost of the protocol while subjected to the physical constraints imposed by the LCD condition (both aspects are elaborated later in this text). The driving profiles that are discovered by the ML scheme, exemplified in Fig.~\ref{fig:omega}, are superior to previously-proposed protocols~\cite{delCampo_2013,Beau_2016}. 
	Furthermore, the ML approach helps uncovering a fundamental problem with a previously-suggested energetic cost metric, which under some conditions violate basic physical principles  (Carnot bound).  
	We show a systematic optimization path based on state-of-the-art ML tools, which permits a search in a large multidimensional variational parameter space;  the space consists of all functions fulfilling STA conditions. The advantage of the DP-ML scheme derives in it using the exact gradients of the quantity of interest with respect to variational parameters, hence reducing the number of required iterations to reach an extremum~\cite{Juan_2020,PRXQuantum.2.020332}.
	
	{\it Quantum Otto refrigerators.--} 
	Prime examples of thermal machines are heat engines and refrigerators~\cite{Abah_2012,PhysRevLett.112.030602,aad6320,Rezek_2006,Rezek_2009}. 
	While the first performs work by utilizing heat current from a hot reservoir, refrigerators extract heat from a cold bath using external work. As a thermodynamic process, refrigerators attain their maximal (Carnot) cooling efficiency, $\epsilon_{\rm C}=\frac{T_1}{T_2-T_1}$ (with $T_{1,2}$ as the temperatures of the cold and hot reservoirs, respectively) for an infinitely-slow (adiabatic) process. Yet, for such processes the power output, defined as the extracted heat over cycle time, $J_c=\frac{\langle Q_4 \rangle}{2 \tau}$, is null due to the infinitely long cycle time, $\tau \to \infty$. 
	For a finite-time cycle, the efficiency decreases, and the power output increases. 
	Therefore, the core question of finite-time thermodynamics is: 
	{\it What is the optimal cycle for a figure of merit given by the cooling efficiency times output power?} 
	
	The quantum Otto refrigerator is depicted in Fig.~\ref{fig:ref_scheme}. We choose a working medium consisting of a harmonic oscillator governed by the time-dependent Hamiltonian~\cite{Abah_2016}
	\be
	H_0(t) = \frac{1}{2m}p^2 + \frac{m \omega(t)^2}{2} x^2.
	\label{eq:H0}\ee
	The cycle consists of an isothermal compression stroke where the frequency $\omega(t)$ increases from $\omega_1$ at $t=0$ to $\omega_2$ at $t=\tau$. Then, the engine thermalizes with a hot bath in an isochoric stroke, followed by an isothermal expansion of duration $\tau$ back to the frequency $\omega_1$, and an isochoric stroke in which heat $\langle Q_4 \rangle$ is extracted from a cold bath. The thermalization strokes are assumed instantaneous.
	
	{\it STA and counterdiabatic driving.--} The goal of the STA driving is to speed up the compression and expansion strokes thus enhance the figure of merit. By adding the nonadiabatic driving $H_{\rm STA}(t)$ to Eq.~(\ref{eq:H0}), the system's final state after a time evolution from $t=0$ to $t=\tau$ exactly matches the outcome of an adiabatic approximation-based time-evolution of $H_0(t)$~\cite{Berry_2009,delCampo_2013}. 
	A further canonical transformation of $H_{\rm STA}(t)$ leads to the
	LCD Hamiltonian of a  harmonic oscillator~\cite{SI,Abah_2018} with frequency $\Omega(t)^2 \equiv \omega(t)^2 - \frac{3 \dot{\omega}(t)^2}{4 \omega(t)^2} + \frac{\Ddot{\omega}(t)}{2\omega(t)}$ \cite{SI}.  This 
	modified driving should fulfil the following conditions~\cite{delCampo_2013}, 
	\bea
	\omega(0) & = & \omega_1, \quad \dot{\omega}(0) = 0, \quad \ddot{\omega}(0) = 0, \nonumber \\ 
	\omega(\tau) & = &  \omega_2, \quad \dot{\omega}(\tau) = 0, \quad \ddot{\omega}(\tau) = 0,
	\label{eq:STAcond}\eea
	which ensure that the final state of the system is identical (phase included) to the state resulting from an adiabatic time evolution of $H_0(t)$.
	
	During compression (AB) and expansion (CD) strokes (Fig.~\ref{fig:ref_scheme}), the system is thermally isolated and work is applied. Using the adiabatic solution of the time-dependent Schr{\"o}dinger equation ~\cite{Husimi_1953,Lohe_2009}, the mean value of work is
	\bea
	\langle W_1\rangle = \frac{\hbar \omega_{2}}{2} \left( 1- \frac{\omega_1}{\omega_2} \right) \coth{\left(\frac{\beta_1 \hbar \omega_1}{2} \right)},  
	\label{eq:W1}\eea
	and similarly for $\langle W_3 \rangle$ by replacing $1 \leftrightarrow 2$. Furthermore, the mean heat extracted during the DA stroke is~\cite{Abah_2016}
	\bea
	\langle Q_4\rangle = \frac{\hbar \omega_{1}}{2} \left[ \coth{\left(\frac{\beta_1 \hbar \omega_1}{2} \right)} - \coth{\left(\frac{\beta_2 \hbar \omega_2}{2} \right)}  \right].  
	\label{eq:Q4}\eea
	We estimate the energetic cost of the STA driving with the time-averaged Schmidt norm of $H_{\rm STA}(t)$ ~\cite{Zheng_2016,Campbell_2017},
	\be
	\langle C_{AB}\rangle = \coth\left(\frac{ \beta_1 \hbar \omega_1 }{2} \right) \frac{\hbar \sqrt{3}}{4\tau} \int_0^\tau dt \frac{1}{\omega(t)}  \left| - \frac{3 \dot{\omega}_t^2}{4 \omega(t)^2} + \frac{\Ddot{\omega}_t}{2\omega(t)} \right|.
	\label{eq:cost}
	\ee
	$\langle C_{CD}\rangle$ is obtained by switching the temperature and frequency $\beta_1, \omega_1$ to $\beta_2, \omega_2$, respectively, see Fig.~\ref{fig:ref_scheme}. 
	More details are included in ~\cite{SI}.
	Below, we show that the energetic cost, Eq.~(\ref{eq:cost}), preserves the physical (Carnot) bound, which is missed by other suggested cost metrics.  
	
	{\it Optimization procedure.--} Our goal is to enhance the figure of merit $\chi$ defined as the product of the cooling efficiency 
	$ \epsilon$ with the heat extracted per cycle, $J_c$,
	\be
	\chi \equiv \epsilon J_c  =  \frac{\langle Q_4 \rangle}{\langle W_1 \rangle + \langle W_3 \rangle + \langle C_{AB}\rangle + \langle C_{CD}\rangle } \times \frac{\langle Q_4 \rangle}{2 \tau}.
	\label{eq:chi}\ee
	Motivated by Ref.~\cite{delCampo_2013}, a possible way to boost the figure of merit could be by using a polynomial ansatz, which by construction satisfies the initial conditions of Eq.~(\ref{eq:STAcond}),
	\bea
	\omega (t) & = & \omega_1 + \Delta \omega \sum_{n=3}^{N_{\rm max}} \alpha_n \left(\frac{t}{\tau}\right)^n.
	\label{eq:poly}\eea
	Here, $\Delta \omega = \omega_2-\omega_1$. 
	A widely used ansatz which satisfies Eq.~(\ref{eq:STAcond}) consists of $\left( \alpha_3,\alpha_4,\alpha_5 \right)=\left(10,-15,6\right)$ and all the other $\alpha=0$, depicted as the dashed-dotted line in Fig.~\ref{fig:omega}. We use it throughout the paper as a benchmark. 
	
	The only quantity that depends on the transient values of $\omega(t)$ is the energetic cost function. Therefore, once the physical parameters ($\beta_i, \omega_i$) are set, the optimal cooling protocol minimizes the energetic cost $\langle C_i\rangle$ [Eq.~(\ref{eq:cost})]. Thus, we devise a cost function that includes $\langle C_i\rangle$, along with penalties for deviating from the STA constraints, Eq.~(\ref{eq:STAcond}). Details are given in \cite{SI}.
	For generality, we represent $\omega(t)$ as a neural network (NN) whose parameters are optimized using automatic differentiation (AD), which allows to compute exact gradients with respect to the NN's parameters. We use Adam~\cite{Adam}, a first-order gradient-based optimization algorithm, to optimize our objective function. This process is performed for a large ensemble of 1000 initial conditions for the NN, out of which the optimal strategies are selected.
	\begin{figure}[h!]
		\includegraphics[width=\columnwidth]{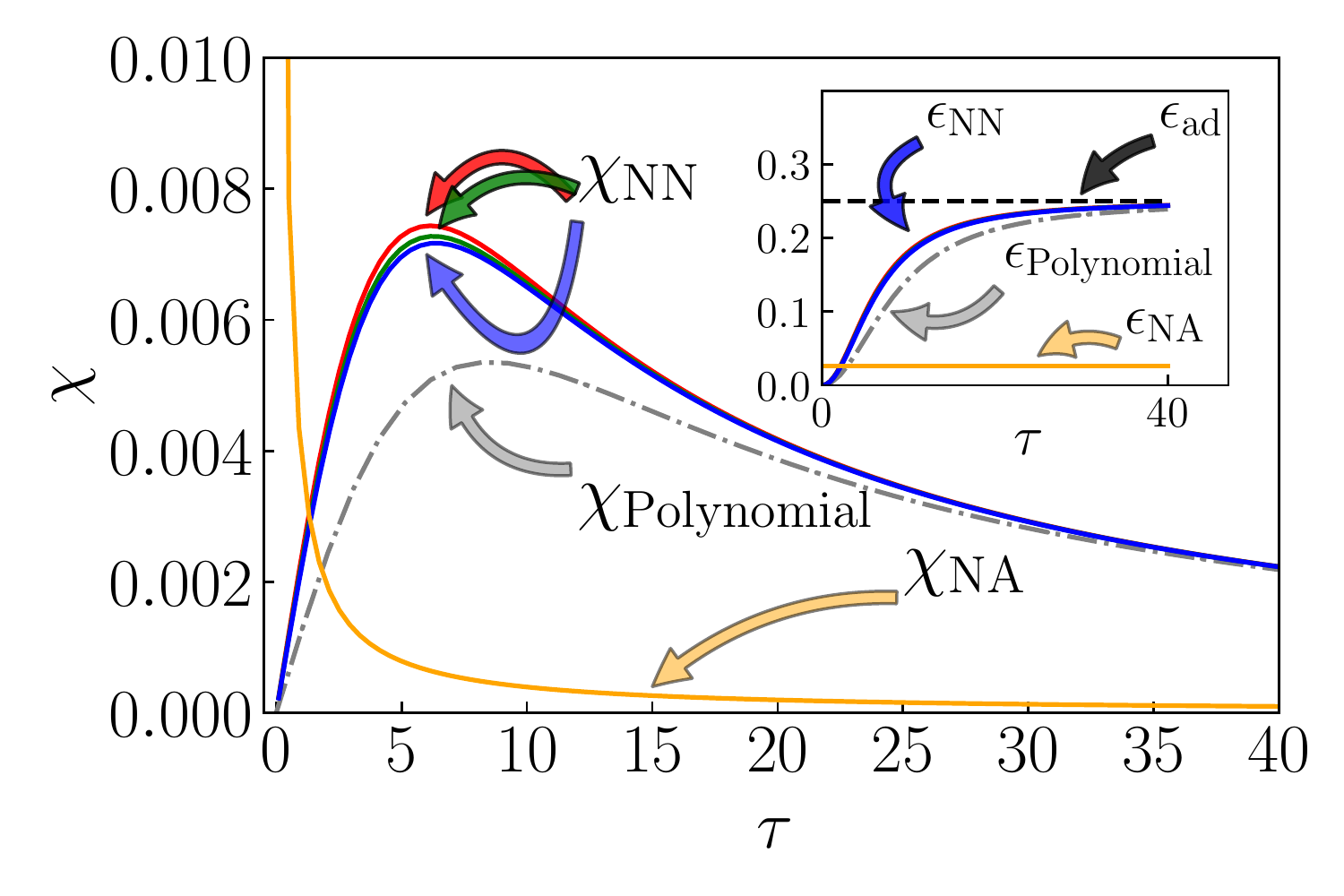} 
		\vspace{-7mm}
		\caption{Figure of merit of the Otto refrigerator $\chi$, Eq.~(\ref{eq:chi}), depicted as a function of the expansion or compression stroke times  $\tau$
			for the frequency ramps $\omega(t)$ of Fig.~\ref{fig:omega}. 
			The gray dashed-dotted line displays the polynomial ansatz, Eq.~(\ref{eq:poly}). Solid lines represent NN results. For comparison, the nonadiabatic sudden-frequency change, $\chi_{\rm NA}$, is plotted in light orange. The NN optimization scheme almost doubles the figure of merit---compared to the polynomial ansatz. The inset shows the corresponding cooling efficiency and the ideal adiabatic limit, $\epsilon_{\rm ad}$. The optimal NN strategies approach the adiabatic efficiency faster than the polynomial benchmark. }
		\label{fig:chi}
	\end{figure}
	
	{\it Results.--} Examples of optimal expansion profiles $\omega(t)$ are depicted in Fig.~\ref{fig:omega};
	the compression stroke is a time-reversed version of it, $\omega(\tau-t)$. 
	We note that during our optimization process, which penalizes for deviations from the initial and final-time conditions, 
	one finds local minima in which the latter are not met. In order to satisfy those, frequency profiles were stretched, in addition to being smoothened in order to become physically realizable~\cite{SI}.  
	In Fig.~\ref{fig:omega}(a), we observe that optimal strategies share a similar feature of a ``late-bloomer", hence they are very different from the polynomial ansatz of Eq.~(\ref{eq:poly}) depicted as a dashed-dotted line in Fig.~\ref{fig:omega}. 
	The inset shows the first and second derivatives of one of these profiles. 
	The two derivatives rise together, which results in the minimization of the energetic cost function as we discuss in the next section. Fig.~\ref{fig:omega}(b) displays the corresponding frequency ramp, $\Omega(t)$, see text above Eq.~(\ref{eq:STAcond}).
	This frequency would have to be followed in order to get the exact same final state as would be achieved by an \emph{adiabatic} driving, which follows $\omega(t)$. 
	
	In Fig.~\ref{fig:chi} we compare the figure of merit $\chi$, Eq.~(\ref{eq:chi}), from the different strategies. Overall, we find an almost two-fold improvement of NN profiles over the benchmark. The peak for the NN-based strategies occurs earlier than for the polynomial, at around $\tau = 6$. 
	Further, we compare these performances to the nonadiabatic, step function strategy~\footnote{We use the nonadiabatic parameter $Q_{\rm NA}^* = \frac{\omega_1^2+\omega_2^2}{2\omega_1 \omega_2}$, which alters Eqs.~(\ref{eq:W1}) and (\ref{eq:Q4}) (see Refs.~\cite{Abah_2012,Abah_2016} for details)}. 
	As expected, the latter strategy could be beneficial for short stroke times, $\tau < 2$, because the energetic cost of maintaining STA is high. Yet, it is inferior at longer cycles. We further plot the cooling efficiency, $\epsilon$, as a function of stroke duration (inset). The black dashed line is the ``ideal" adiabatic efficiency  $\epsilon_{\rm ad}$, where the heat $\langle Q_4 \rangle$ 
	and work $\langle W_1 \rangle$, $\langle W_3 \rangle$ attain their adiabatic values and the associated driving energetic cost is neglected (note that the relation to the Carnot cooling efficiency is $\epsilon_C \geq \epsilon_{\rm ad}$). NN-based strategies are able to obtain a higher efficiency, and, in turn, similarly to the polynomial benchmark, they approach the adiabatic limit at large $\tau$. 
	
	\begin{figure}[h!]
		\includegraphics[width=0.8\columnwidth]{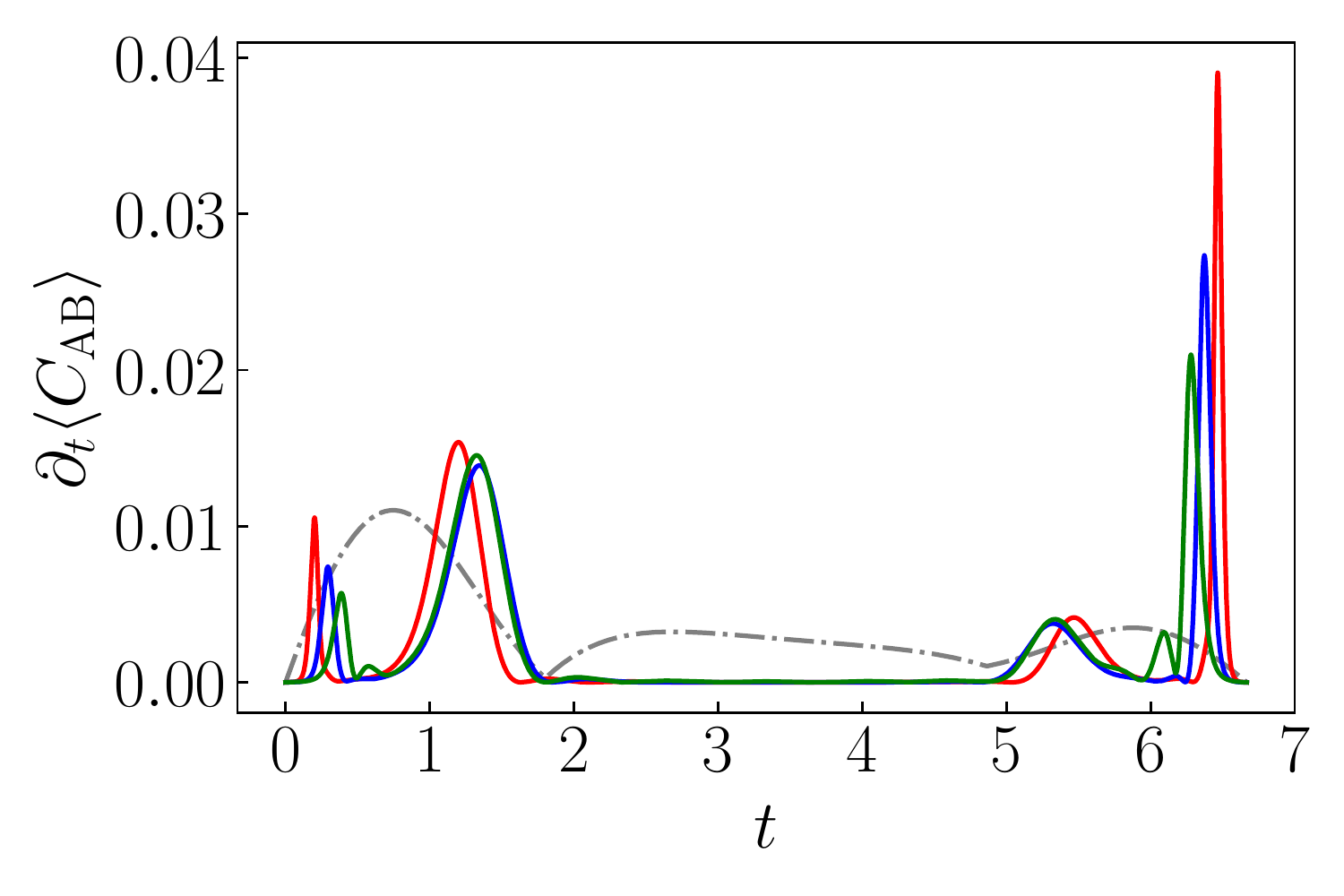}
		\vspace{-7mm}
		\caption{Mean instantaneous energetic cost of the STA drive as a function of time, $\partial_t \langle C\rangle$,  defined by the Schmidt norm of $H_{\rm STA}(t)$,  Eq.~(\ref{eq:cost}). 
			The dashed line corresponds to the polynomial ansatz, Eq.~(\ref{eq:poly}). Solid lines correspond to the 
			optimized NN of  Figs.~\ref{fig:omega} and~\ref{fig:chi} with
			$\tau = 6$. } 
		\label{fig:d_tC}
	\end{figure}
	The cooling protocols minimize the cost metric. 
	In Fig.~\ref{fig:d_tC}, we show the mean instantaneous energetic cost of the STA drive as a function of time, $\partial_t \langle C_{AB}\rangle $. 
	The NN optimization is able to reduce the energetic cost by a factor of two, compared to the polynomial ansatz.
	Recall that in Eq.~(\ref{eq:cost}) the first and second derivative terms appear with opposite signs. 
	Hence, in order to reduce the energetic cost one should envisage functions in which both derivatives follow similar temporal features. This is where the power of NN-based optimization techniques come to play. The DP-ML method greatly reduces the {\it instantaneous} cost by realizing  functions with this property (see inset in Fig.~\ref{fig:omega}(a)). 
	
	Next, we discuss different energetic cost metrics and their adherence to thermodynamical principles. 
	While  we employed Eq. (\ref{eq:cost}), previous studies suggested the time-average of the mean STA driving, $\langle H_{\rm STA} \rangle = 1/\tau \int_0^\tau dt \langle H_{\rm STA}(t) \rangle$, as the energetic cost of STA~\cite{Abah_2017,Abah_2019,Abah_2020}.
	We show that this expression can lead to unphysical results: The system is a refrigerator,
	$\langle Q_4 \rangle > 0$, yet the total work plus associated STA cost are negative,
	$\langle W_1 + W_3 \rangle+\langle H_{\rm STA} \rangle <0$, thus yielding a negative efficiency.
	
\begin{figure}[h!]
		\includegraphics[width=\columnwidth]{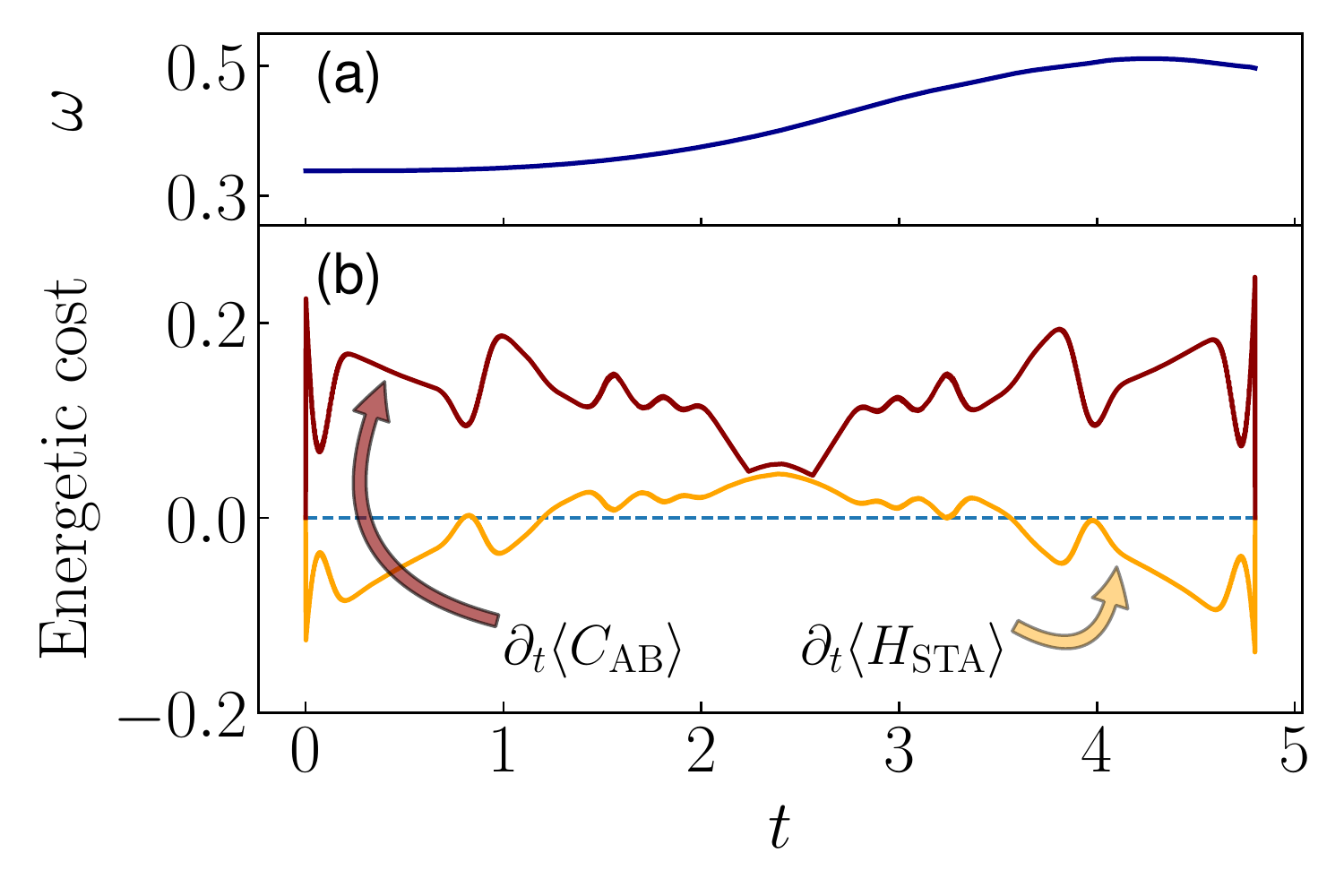} 
		\vspace{-7mm}
		\caption{Exemplifying the failure of the energetic cost definition $\langle H_{\rm STA}\rangle$.
			(a) Frequency ramp profile, $\omega(t)$,  optimized to minimize $\langle H_{\rm STA}\rangle$.
			The initial frequency $\omega_1$ is set at $0.34$ such that the sum of the average work needed for the expansion and compression strokes, $\langle W_1 + W_3 \rangle$, is relatively small.  
			(b) Instantaneous energetic cost  as a function of time,
			$\partial_t \langle C_{AB}\rangle$ of Eq.~(\ref{eq:cost}) (light line),
			and  $\langle H_{\rm STA} \rangle$,  which is based on the time-averaged mean STA driving
			(dark line). For the latter, the overall input work plus energetic cost is negative,  hence unphysical. In contrast, using $\langle C(t)\rangle$ 
			as the energetic cost yields  the cooling efficiency $\epsilon = 0.47 < \epsilon_{\rm ad} \leq \epsilon_C=3$; 
			$\tau = 4.8$,  $\omega_1=0.34$, $\omega_2=0.5$, 
			$\beta_1=1$,  $\beta_2=0.75$.}
		\label{fig:Costs}
	\end{figure}

	We compare cost metrics in Fig.~\ref{fig:Costs}. We use parameters close to the edge of the cooling window
	(determined by the condition $\beta_1 \omega_1 < \beta_2 \omega_2$, see Eq.  (\ref{eq:Q4})). This choice leads to relatively small values of $\langle Q_4 \rangle$ and $\langle W_1 + W_3 \rangle$,
	compared with parameters used in Fig. \ref{fig:omega}, and it allows us to demonstrate
	the incentive for devising a different energetic cost function for STA protocols. NN optimization with the cost metric $\langle H_{\rm STA} \rangle $ yields  profiles that allow cooling, see Fig.~\ref{eq:cost} (a) for an example, yet give an overall {\it negative} energetic cost;
	in Fig.~\ref{eq:cost}(b)  we show the instantaneous contribution $\partial_t \langle H_{\rm STA}\rangle$, which is mostly negative. In contrast, the metric $\langle C\rangle$ of Eq. (\ref{eq:cost}) is positive throughout. The cooling efficiency for these parameters becomes negative for the $\langle H_{\rm STA} \rangle$ energetic cost---which is unphysical:
	By fine-tuning parameters, while maintaining the cooling condition, one can achieve an efficiency that exceeds Carnot ~\footnote{Criticism to this approach, albeit from a different aspect, were raised in Ref.~\cite{Kosloff_2014}}. In contrast,  the Schmidt norm-based definition  recovers a positive value for the overall energetic cost and an efficiency $\epsilon \leq \epsilon_C$, in compliance with thermodynamical laws.
	
	{\it Discussion.--}
	We demonstrated the potential and advantage of NN combined with AD in the field of quantum control of finite-time thermodynamics.
	Our method allowed to discover driving protocols of the strokes of an Otto engine that perform twice as better compared to previously-conceived solutions. This scheme was able to find a nontrivial family of functions in which the first and second derivatives follow each other; from our results we conclude that it is a crucial property of the cost function. In \cite{SI} we discuss an attempt to optimize a simpler ansatz, which can mimic the NN's ``late blooming" strategy. Our conclusion is that NN-based results are difficult to generate with a simple analytical form.  
	Furthermore, employing the DP-ML optimization scheme enabled us to uncover a flaw in a previous  definition for the cost of STA driving. In contrast, our modified definition,provided physical results, which obeyed the Carnot bound on one hand, and reached the adiabatic limit on the other hand. We point out that among the plethora of energetic cost metrics suggested in the literature~\cite{DelCampo_2014,Funo_2017,Abah_2019} we do not aim here to find which one is the most appropriate. However, ML optimization naturally identified violations to thermodynamical laws. Since our optimization method is general, it can be easily turned to optimize other cost functions and figures of merits with little effort.  
	
	Our framework could be directly applied in other control problems, such as entropy reduction in closed systems ~\cite{Sgroi_2021}, dynamical decoherence control~\cite{Porotti_2019}, 
	steering chemical reactions~\cite{de2019synthetic}, and for the design of quantum electronic and thermal machines~\cite{Santos_2015}. We paved the way for solving hard-constrained problems using  state-of-the-art ML tools, by orchestrating an objective as the minimum of a cost function. More generally, our study shows that ML has an advantage over standard theoretical tools in designing quantum devices, thus making them favorable for an experimental realization. 
	
	{\it Acknowledgments.--}
	We acknowledge fruitful discussions with Adolfo del Campo, Obinna Abah, Rodrigo A. Vargas-Hernández, Junjie Liu and Aharon Brodutch.
	The work of IK was supported by the Centre for Quantum Information and Quantum Control (CQIQC) at the University of Toronto. JC acknowledges support from the Natural Sciences and Engineering Research Council of Canada (NSERC), the Shared Hierarchical Academic Research Computing Network (SHARCNET), Compute Canada, Google Quantum Research Award, and the Canadian Institute for Advanced Research (CIFAR) AI chair program, and companies sponsoring the Vector Institute.
	DS acknowledges support from an NSERC Discovery Grant and the Canada Research Chair program.

	\pagebreak
	\widetext
	\begin{center}
		\textbf{Supplemental Material: Optimal control of quantum thermal machines using machine learning}
	\end{center}
	
	\renewcommand\thefigure{S\arabic{figure}}
	\renewcommand{\theequation}{S\arabic{equation}}
	\setcounter{figure}{0}
	\setcounter{equation}{0}
	
	\appendix
	\section{Derivation of the energetic cost proxy}
	For completeness we derive the modified Hamiltonian under local counterdiabatic (LCD) driving bellow.
	The original driven Hamiltonian, $H_0(t)$, and the counterdiabatic term, $H_{CD}(t)$, are given explicitly by
	\bea
	H_0(t)&=&\frac{p^2}{2m} + \frac{m x^2 \omega(t)^2}{2},
	\nonumber\\
	H_{CD}(t)&=&-\frac{\dot{\omega(t)}}{4\omega(t)}\left(xp + px\right).
	\eea
	The LCD Hamiltonian is
	\bea
	H_{\rm LCD}(t) & = & U_x^\dagger \left(H_0(t) + H_{\rm CD}(t) -i \hbar \dot{U}_x U_x^\dagger \right)U_x \nonumber \\ 
	& = & U_x^\dagger \left[\frac{p^2}{2m} + \frac{m x^2 \omega(t)^2}{2} -\frac{\dot{\omega(t)}}{4\omega(t)}\left(xp + px\right) \right.  \left. + \frac{m x^2}{4\omega(t)} \left(\Ddot{\omega}_t - \frac{\dot{\omega}_t^2}{\omega(t)} \right) \right] U_x,
	\eea
	where $U_x=\me^{i\frac{m x^2 \dot{\omega}_t}{4\hbar \omega(t)}}$ is a time-dependent, local operator which is in charge of eliminating the $x,p$ coupling due to the counterdiabatic term $H_{\rm CD}(t)$. Applying $U_x$ leads to 
	\bea
	H_{\rm LCD}(t) & = & H_0(t) + H_{\rm STA}(t) \nonumber \\
	& = & \frac{p^2}{2m} + \frac{mx^2}{2}\left( \omega(t)^2 - \frac{3 \dot{\omega}_t^2}{4 \omega(t)^2} + \frac{\Ddot{\omega}_t}{2\omega(t)} \right) \nonumber \\
	& = & \frac{p^2}{2m} + \frac{mx^2}{2}\Omega(t)^2.
	\eea
	Here, $\Omega(t)^2 \equiv \omega(t)^2 - \frac{3 \dot{\omega}_t^2}{4 \omega(t)^2} + \frac{\Ddot{\omega}_t}{2\omega(t)}$.
	Naturally, a trap inversion condition follows our definition of $\Omega \ge 0$ for $t \in [0,\tau]$, with compression or expansion stroke time $\tau$. 
	
	Following Refs.~\cite{Zheng_2016,Campbell_2017} we utilize Schmidt norm of the additional STA term, 
	$H_{\rm STA}(t)=H_{\rm LCD}(t) -H_{0}(t)$ 
	to serve as a proxy to the energetic cost of the LCD drive, $\langle C\rangle$.
	\bea
	\langle C\rangle & \equiv & \frac{1}{\tau}\int_0^\tau dt \|H_{\rm STA}(t) \| \nonumber \\
	& = & \frac{m}{2} \frac{1}{\tau}\int_0^\tau dt \sqrt{\left\langle x^4 \right\rangle_t^{\rm LCD}} \left| - \frac{3 \dot{\omega}_t^2}{4 \omega(t)^2} + \frac{\Ddot{\omega}_t}{2\omega(t)} \right|.
	\label{eq:C}
	\eea
	For an eigenstate of a harmonic oscillator, $\ket{n}$, $\braket{n}{x^4}{n} = \frac{\hbar^2}{4 m^2 \omega^2}\left( 6n^2 + 6n +3 \right)$. Therefore, for a harmonic oscillator in a canonical thermal state given by $\rho_{th} = \sum_{n=0}^\infty p_n\ket{n}\bra{n}$ with $p_n = \frac{\me^{-\beta E_n}}{\mathcal{Z}}$,
	\be
	{\rm Tr} [\rho_{th}x^4]
	= \frac{3 \hbar^2}{4 m^2 \omega^2} \left[\coth\left(\frac{ \beta \hbar \omega }{2}\right)\right]^2.
	\label{eq:x4exp}\ee
	Following Ref.~\cite{Beau_2016} we can now evaluate this expectation value for the LCD state,
	\bea
	\braket{\Psi^{\rm LCD}_{x,t}}{x^4}{\Psi^{\rm LCD}_{x,t}} 
	& = & \braket{\Psi^{\rm CD}_{x,t}} {\me^{-i\frac{m x^2 \dot{b}_{\rm ad}}{2\hbar b_{\rm ad}}} x^4 \me^{i\frac{m x^2 \dot{b}_{\rm ad}}{2\hbar b_{\rm ad}}}} { \Psi^{\rm CD}_{x,t}}  \nonumber \\
	& = & \braket{\Psi^{\rm CD}_{x,t}} {x^4} { \Psi^{\rm CD}_{x,t}} \nonumber \\
	& = & \frac{1}{b_{\rm ad}}\braket{\Psi_{\frac{x}{b_{\rm ad}},t=0}} {x^4} { \Psi_{\frac{x}{b_{\rm ad}},t=0}} \nonumber \\
	&=&b_{\rm ad}^4\frac{\hbar^2}{4m^2\omega_i^2}\left(6n^2+6n+3\right).
	\eea
	Averaging with respect to the thermal state we get
	\bea
	\left\langle x^4 \right\rangle_t^{\rm LCD} =  b_{\rm ad}^4 \frac{3 \hbar^2}{4 m^2 \omega_i^2} \left[\coth\left(\frac{ \beta_i \hbar \omega_i }{2}\right)\right]^2,
	\eea
	where we used Eq.~(\ref{eq:x4exp}) in the last step, and with $b_{\rm ad}(t) = \sqrt{\omega_i/\omega(t)}$; $\omega_i$ and $\beta_i$ are the initial-time frequency and inverse temperatures (depending on the stroke). Plugging this expression back into Eq.~(\ref{eq:C}) yields the expression for the energetic cost,
	\bea
	\langle C_i \rangle & = & \frac{\hbar \sqrt{3}}{4} \frac{1}{\tau}\int_0^\tau dt \frac{ b_{\rm ad}^2 }{ \omega_i} \coth\left(\frac{ \beta_i \hbar \omega_i }{2} \right) \left| - \frac{3 \dot{\omega}_t^2}{4 \omega(t)^2} + \frac{\ddot{\omega}_t}{2\omega(t)} \right| \nonumber \\
	& = & \coth\left(\frac{ \beta_i \hbar \omega_i }{2} \right) \frac{\hbar \sqrt{3}}{4\tau} \times \int_0^\tau dt \frac{1}{\omega(t)}  \left| - \frac{3 \dot{\omega}_t^2}{4 \omega(t)^2} + \frac{\Ddot{\omega}_t}{2\omega(t)} \right|.
	\eea
	The total energetic cost of the STA  is the sum of two terms corresponding to the two strokes: $\langle C \rangle=\langle C_{\rm AB} \rangle + \langle C_{\rm CD} \rangle$.
	
	\begin{figure}
		\includegraphics[scale=0.7]{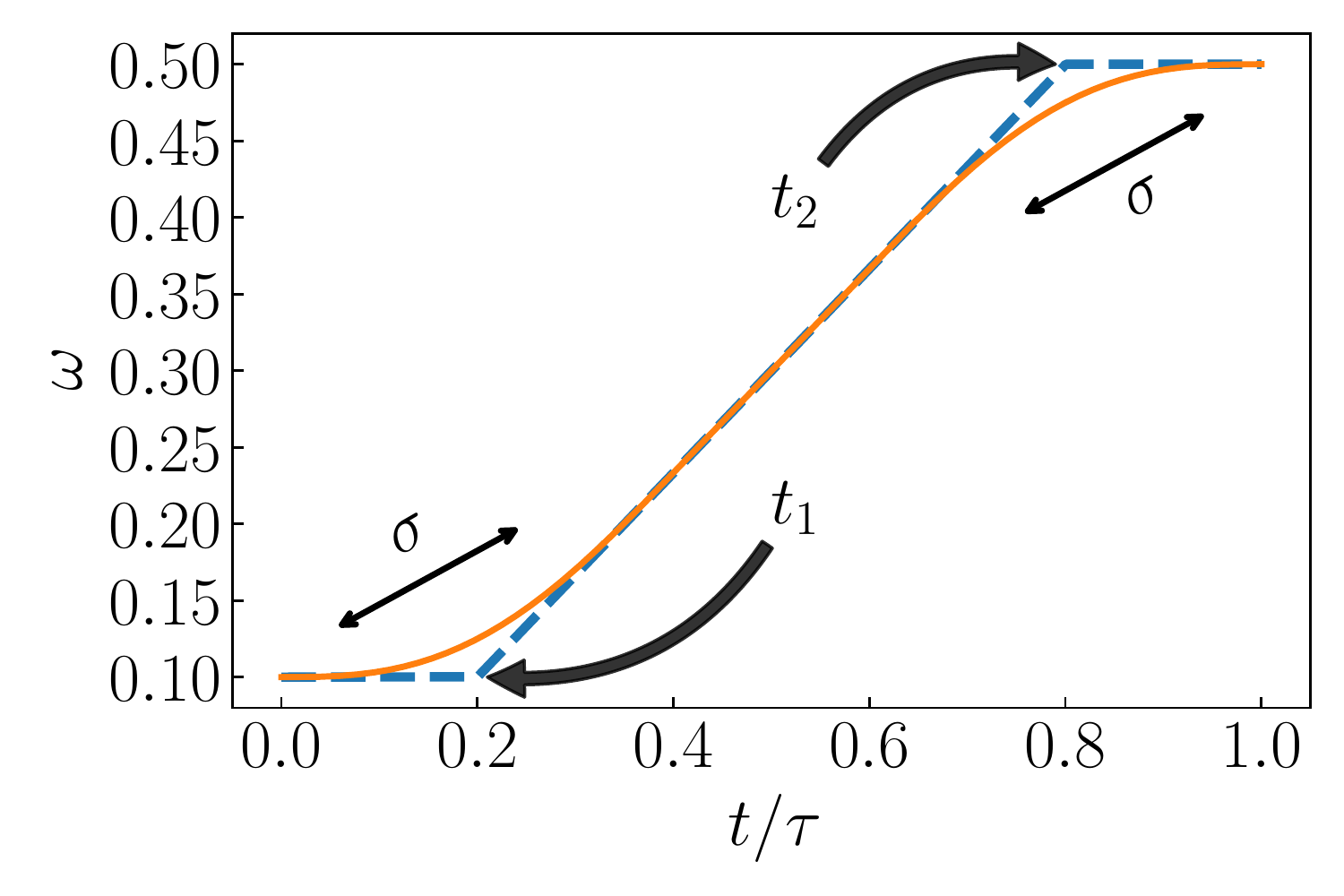} 
		\caption{A schematic of a different ansatz. We display the bare function, which contains discontinuities in its derivatives (dashed-blue)
			and its smoothed interpolated counterpart (full-orange). This ansatz utilizes a mostly flat region where both derivatives of $\omega(t)$ vanish, separated by a linear ramp in the time domain $t \in [t_1, t_2]$. The smoothed ansatz contains two regions of width $\sigma$ around $t=t_1, t_2$. At the edges of these regions  the smoothened function and its derivatives match the original (dashed) function.}
		\label{fig:linearansatz}
	\end{figure}
	
	\section{Additional ansatz}
	By examining the STA energetic cost function in Eq.~(\ref{eq:C}), one might suggest the following ansatz for minimizing the cost, which is depicted in Fig.~\ref{fig:linearansatz}. This function  is made of a mostly flat region where both derivatives of $\omega(t)$ vanish. These domains are separated by a linear ramp at times $t \in [t_1, t_2]$. This kind of function is shown in dashed-blue. In order to avoid discontinuities in the derivatives of $\omega(t)$ one can smooth this function at some times of range $\sigma$, fixing the values of the interpolated function and its two derivatives on both edges of the region to match those of the original function. This interpolation is solved by a set of linear equations, and is shown in full line. Note that this type of function allows for a ``late bloomer" strategy that the Neural Network (NN) method discovered as optimal.
	
	Next, we turn to find the optimal values of this suggested function in terms of the energetic cost function by examining the three dimensional space $(t_1, t_2, \sigma)$. We find that the optimal solution for this ansatz is very similar to the polynomial ansatz (main text). Although the present family of functions allows for different profiles than the polynomial ansatz, it is limited to a set of somewhat trivial functions, which cannot fully minimize the integrand of the cost function. As a result, these functions do not allow for a behavior as well-captured by the NN, i.e. a function in which both derivatives follow each other. In fact, using this ansatz and mimicking a ``late blooming" strategy with a large $t_1$ is energetically disadvantageous.   
	
	\section{Neural network optimization}
	Our NN is has a single input of time, $t$. It is passed through a ``polynomial layer", in which a polynomial in the form of Eq. (7) (Main text) with $N_{\rm max}=10$ is generated. This function is fed into three layers of $100$ neurons with a sigmoid activation function followed by one output neuron with a sigmoid activation function. The intermediate, polynomial basis layer was found to assist in reaching a faster convergence, with practically no dependence on $N_{\rm max}$. We apply our optimization scheme onto the following cost function 
	\bea
	CF(\theta) & = & \langle C_{\rm AB}  \left( \omega(t), \dot{\omega}(t), \ddot{\omega}(t) \right) \rangle \nn \\
	& + &  P_{\omega(0)} \left| \omega(0) - \omega_1 \right| + P_{\dot{\omega}(0)}\left| \dot{\omega}(0) \right| \nn \\
	& + & P_{\ddot{\omega}(0)}\left| \ddot{\omega}(0) \right| + P_{\omega(\tau)}\left| \omega(\tau) - \omega_2 \right|  \nn \\
	& + &  P_{\dot{\omega}(\tau)}\left| \dot{\omega}(\tau) \right| + P_{\ddot{\omega}}(\tau)\left| \ddot{\omega}(\tau) \right| \nn \\
	& + & P_{\Delta \omega}~{\rm ReLU}(\omega(0)-\omega(\tau)),
	\label{eq:CF}\eea
	where $\theta$ are the NN parameters, and ${\rm ReLU}$ is a rectified linear unit. The first term in the above is the STA energetic cost, appearing in Eq.~(\ref{eq:C}) for the expansion cycle AB (See Fig.~1 Main text). 
	The next six terms correspond to the six STA requirements in Eq.~(2) of the Main text. 
	The last term in Eq. (\ref{eq:CF}) penalizes a situation
	in which $\omega(\tau) - \omega(0) < 0$ that we found to occur sometimes. 
	
	We start with random initial network parameters, $\theta$, and run the first (out of four) stochastic gradient descent passes of 1000 steps with a large value of $P_{\Delta \omega}$ and relatively low values for the other $P$s. We choose the parameters that yield the lowest cost function value along the pass, and use those for the next step. For the next passes, we gradually increase the values of all the $P$s but $P_{\Delta \omega}$, which is annualized for the final pass. 	
\end{document}